\documentclass[10pt,journal,compsoc]{IEEEtran}

\usepackage{times}
\usepackage{epsfig}
\usepackage{amsmath}
\usepackage{amssymb}
\usepackage{multirow}
\usepackage{graphicx}
\usepackage{arydshln}
\usepackage{balance}
\usepackage{hhline}

\usepackage{color, colortbl}
\definecolor{min}{rgb}{1,0.4,0.2}
\definecolor{max}{rgb}{0.4,0.7,0.9}
\definecolor{Gray}{gray}{0.8}

\ifCLASSOPTIONcompsoc
  \usepackage[nocompress]{cite}
\else
  \usepackage{cite}
\fi

\ifCLASSINFOpdf

\else
  
\fi

\begin{document}

\title{Fast Free-text Authentication via Instance-based Keystroke Dynamics}

\author{Blaine~Ayotte,~\IEEEmembership{Student~Member,~IEEE,}
        Mahesh~Banavar,~\IEEEmembership{Senior~Member,~IEEE,}
        Daqing~Hou,~\IEEEmembership{Member,~IEEE}
        and~Stephanie~Schuckers,~\IEEEmembership{Senior~Member,~IEEE}
\IEEEcompsocitemizethanks{\IEEEcompsocthanksitem Department of Electrical and Computer Engineering, Clarkson University, Potsdam,
NY, 13676 USA.\protect\\
E-mails: \{ayottebj, mbanavar, dhou, sschucke\}@clarkson.edu
}


}

\markboth{Journal of \LaTeX\ Class Files,~Vol.~14, No.~8, August~2015}%
{Shell \MakeLowercase{\textit{et al.}}: Bare Advanced Demo of IEEEtran.cls for IEEE Biometrics Council Journals}

\IEEEtitleabstractindextext{%
\begin{abstract}
Keystroke dynamics study the way in which users input text via their keyboards. Having the ability to differentiate users, typing behaviors can unobtrusively form a component of a behavioral biometric recognition system to improve existing account security. Keystroke dynamics systems on free-text data have previously required 500 or more characters to achieve reasonable performance. In this paper, we propose a novel instance-based graph comparison algorithm called the instance-based tail area density (ITAD) metric to reduce the number of keystrokes required to authenticate users. Additionally, commonly used features in the keystroke dynamics literature, such as monographs and digraphs, are all found to be useful in informing who is typing. The usefulness of these features for authentication is determined using a random forest classifier and validated across two publicly available datasets. Scores from the individual features are fused to form a single matching score. With the fused matching score and our ITAD metric, we achieve equal error rates (EERs) for 100 and 200 testing digraphs of 9.7\% and 7.8\% for the Clarkson II dataset, improving upon state-of-the-art of 35.3\% and 15.3\%.
\end{abstract}

\begin{IEEEkeywords}
Keystroke dynamics, biometrics, continuous authentication, monograph, digraph, fusion, classification
\end{IEEEkeywords}}

\maketitle

\IEEEdisplaynontitleabstractindextext

\IEEEpeerreviewmaketitle

\ifCLASSOPTIONcompsoc
\IEEEraisesectionheading{\section{Introduction}\label{sec:introduction}}
\else
\section{Introduction}
\label{section:introduction}
\fi

\IEEEPARstart{I}{n} today's world, with tremendous amounts of sensitive data becoming digitized, protecting private user data is paramount. Passwords or other single-sign-on (SSO) security measures are typically the only line of defense against attackers. With the large number of accounts and passwords people are expected to remember, people tend to choose easily guessable passwords. For example, it was found in a major data breach that 60\% of passwords were easily guessable \cite{passwords}. An additional form of verification to supplement SSO security schemes is needed to monitor the user of a device to ensure they are authorized. 

Keystroke dynamics is a behavioral biometric that offers strong performance distinguishing users based on typing patterns \cite{teh2013survey,alsultan2013keystroke,banerjee2012biometric}. Keystroke dynamics can be used to provide an additional continuous layer of security to supplement an existing system to detect intruders in a more robust fashion. Furthermore, as most computers already have a keyboard, this layer of continuous security does not require any additional hardware. 

\begin{figure}[htb]
\begin{minipage}[b]{1.0\linewidth}
  \centering
  \centerline{\includegraphics[width=8.0cm]{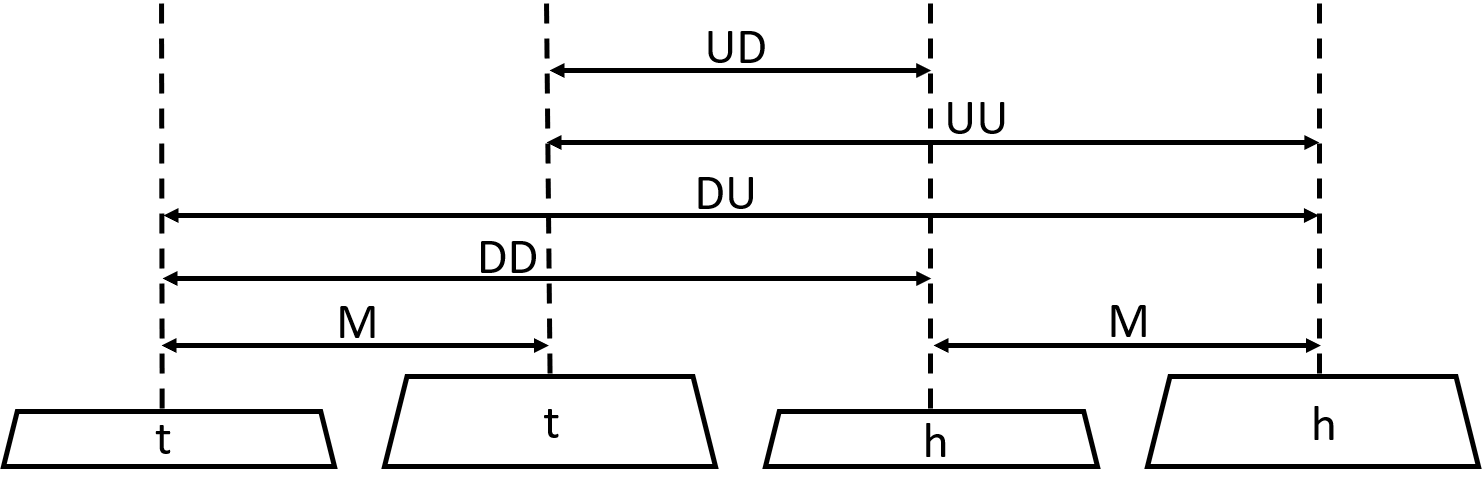}}
  \caption{Graphical representation of how monograph and the four different digraph features can be extracted from two consecutive keystrokes. These digraphs can also be referred to as press-press, press-release, release-press, and release-release (see \cite{harilal2017twos,biosig}).}
  \label{figure:features}
\end{minipage}
\end{figure}

Keystroke dynamics systems have two steps, training and testing. During the training phase, as many keystrokes as feasibly and practically possible are collected from the authorized user and used to build a profile. In many systems, features such as durations of monographs and digraphs (hold time and flight time of key-presses associated with specific letter combinations as shown in Figure \ref{figure:features}) are extracted from the keystrokes. The testing phase consists of keystrokes from an unknown user which are compared to an authorized user's profile to determine if the keystrokes came from the authorized user or an imposter.

There are two main types of keystroke dynamics: fixed-text and free-text. Fixed-text requires the keystrokes of the test sample to exactly match with the keystrokes of the profile. The fixed text keystrokes can constitute a password or any other phrase. Most of the literature for keystroke dynamics is related to fixed-text and performance can be strong on passwords or phrases of around 10 characters. For example, Kilhourhy and Maxion achieved an EER of 9.6\% for a fixed-text sample of ``.tie5Roanl'' \cite{killourhy2009comparing}.

Free-text, on the other hand, puts no restrictions on the keystrokes users can type. Some studies provide guidance on what users should write about, which is considered to be \textit{controlled} free-text. 
In contrast, \textit{uncontrolled} free-text puts absolutely no restrictions on what users can type, capturing user behavior while they naturally type. Continuous authentication in the uncontrolled free-text setting is difficult because users can participate in many different activities while typing. It is possible that a user's typing behavior can vary across activity or content. Additionally, getting enough similar characters in the profile and test samples can be challenging if the user is typing in different contexts. Huang \textit{et al.}, performed a benchmarking study where three leading algorithms were compared across four publicly available datasets and found that algorithms with the same profile and test sample sizes perform consistently worse in the uncontrolled free-text environment \cite{huang2017benchmarking}.

Table \ref{table:intro_table} shows the number of keystrokes for commonly typed texts \cite{cvpr2019}. Requiring large numbers of keystrokes before these systems can detect an intruder could allow for considerable damage to be done. Existing keystroke dynamics research in the uncontrolled free-text environment have been primarily done with large keystroke samples \cite{GP05,idrus2014soft,huang2017benchmarking,sun2016shared}. Many existing algorithms are distribution-based and rely on comparing distributions of digraph durations between the reference and the test user \cite{alsultan2013keystroke,huang2017benchmarking}. Distribution-based algorithms require large numbers of keystrokes for both training and testing, resulting in less frequent authentication or authentication after a significant amount of typing has occurred. Currently, the existing algorithms require 500, or more keystrokes to authenticate users. For example, previous work achieved best EERs on the Clarkson II and Buffalo datasets of 3.6\% and 1.95\% for test sample sizes of 1,000 DD digraphs and 1,000 keystrokes respectively \cite{cvpr2019,huang2017benchmarking}.

\begin{table}[htb]
\centering
\caption{Estimates of character counts for various types of texts \cite{cvpr2019}.}
\label{table:intro_table}
\begin{tabular}{|l|c|c|c|}
\hline
Typed Text & Characters \\
\hline
Average tweet length & 60-70\\
Average sentence & 75-100\\
Phishing email & 120 \\
Average Facebook post & 155 \\
\hline
Gettysburg Address & 1450 \\
Nigerian prince emails &  1500-2500\\
\hline
\end{tabular}
\end{table}

For a continuous authentication system to be useful, users should be authenticated as quickly and often as possible. This will increase overall usability and lead to the acceptability of keystroke dynamics as a behavioral biometric. To increase the speed of authentication, decisions need to be made after as few keystrokes as possible. To reduce the number of keystrokes needed for authentication, we use instance-based algorithms, which compare graph times from the test sample \textit{individually} to the reference profile. Instance-based methods are not foreign to keystroke dynamics, but are extensively used for fixed-text \cite{killourhy2009comparing, zhong2012keystroke}.

In this paper, we propose a novel instance-based metric called the instance-based tail area density (ITAD) metric. The performance of the ITAD metric is compared to algorithms previously used for keystroke dynamics including Manhattan, scaled Manhattan, Mahalanobis, transformed Mahalanobis, and KDE \cite{killourhy2009comparing,biosig,huang2017benchmarking}. Furthermore, the importance of monographs and digraphs, commonly used features in keystroke dynamics, is determined for user authentication. The features are fused at the score level into a single fused matching score using the feature importances determined from a random forest classifier.
The effectiveness of our fused matching score is demonstrated on two publicly available datasets, the Clarkson II \cite{DBLPconficb2017} and Buffalo \cite{sun2016shared} datasets. 
Authentication with as few keystrokes as possible allows imposters to be detected faster and thus better protects sensitive user data. 

The rest of this paper is organized as follows: Section \ref{section:related_work} related work, Section \ref{section:features} the features commonly used in keystroke dynamics research, Section \ref{section:datasets} discusses the two free-text datasets, and Section \ref{section:algorithms} the algorithms used. The algorithms are compared in Section \ref{section:comparing_algs}, and Section \ref{section:feature_selection} discusses how the keystroke features are fused at the score level. Lastly, the ITAD metric is evaluated in Section \ref{section:results}. Concluding remarks and future work are presented in Section \ref{section:conclusion}.

\section{Related Work}
\label{section:related_work}

The amount of research done with free-text is much smaller compared to fixed-text. Furthermore, the majority of free-text research is controlled.  
A survey citing more than 180 works in keystroke dynamics \cite{teh2013survey} finds that as of 2013 there are roughly 8 times more papers working with fixed-text authentication than with free-text authentication.
Of these free-text papers, even fewer are in the uncontrolled setting.

Features used in keystroke dynamics consist primarily of monographs (hold time of a key) and digraphs (elapsed time between two consecutive key-presses) \cite{teh2013survey,monrose1997authentication,GP05,alsultan2013keystroke,umphress1985identity,huang2017benchmarking}. Other common features including trigraphs (elapsed time between three consecutive key-presses) and larger $n$-graphs (elapsed time between $n$ consecutive key-presses) have also been shown to be effective at distinguishing between users \cite{sim2007digraphs}. However, it is unlikely to get enough of these large $n$-graphs that are shared between the profile and test set especially when authenticating users with minimal keystrokes. Other features used in keystroke dynamics consist of relative timing between graphs \cite{GP05}, pressure (only when using specialized keyboards) \cite{allen2010analysis}, splitting the keyboard into regions \cite{park2010user}, typing speed (words per minute) \cite{hempstalk2008one}, error rate of typing \cite{hempstalk2008one,curtin2006keystroke}, press/release ordering \cite{hempstalk2008one}, and percentage of special characters \cite{curtin2006keystroke}. 

Researchers that use both monographs and digraphs typically weight the features equally, i.e. monographs and digraphs contribute equally to the final decision. Joyce and Gupta created a feature vector from keystrokes which included both monographs and digraphs \cite{joyce1990identity}. To perform authentication, the Manhattan distance was used to create a distance score between the profile and test sample. In their experiment the monographs and digraphs carried the same weight. Monrose and Rubin also used monographs and digraphs and weighted them equally \cite{monrose1997authentication}. 

Authentication via free-text keystroke dynamics has been known to work well with large numbers of keystrokes \cite{GP05}. For their research, Gunetti and Picardi evaluated performance on free-text keystroke dynamics systems with 700 to 900 characters. While their smallest test sample was roughly 200 characters, we cannot perform a meaningful comparison to \cite{GP05} because they do not present a ROC or DET curve of their results, but instead a single point. More recently, researchers have started trying to reduce the number of keystrokes required for authentication \cite{bours2015continuous,cvpr2019,biosig}. 

Ayotte \textit{et al.} investigate the performance of existing state-of-the-art free-text distribution-based algorithms on small and large test sample sizes \cite{cvpr2019}. 
The authors experimented with test sample sizes of 100, 200, 500, and 1,000 DD digraphs with a profile size of 10,000 DD digraphs. It was found that by fusing three algorithms (KDE, Energy, and KS \cite{cvpr2019}) at the decision level, comparable performance could be achieved with 500 testing digraphs instead of 1,000. However, for less than 500 digraphs it was found that there was not always enough of the same digraph in the test sample to make a comparison and the performance degraded (EERs could not even be calculated at less than 80 DD digraphs). The authors concluded that the performance of their distribution-based classifier was strong for large test sample sizes (500 or more DD digraphs), but did not perform well with small test samples sizes (200 or fewer DD digraphs).

In a later work by Ayotte \textit{et al.} \cite{biosig}, instance-based algorithms were applied to uncontrolled free-text. Additionally, to capture more information about users, monographs and the four digraphs seen in Figure \ref{figure:features} were used instead of just the DD digraph. This approach was applied to the Clarkson II uncontrolled free-text dataset, and with 30,000 DD digraphs in the profile, achieved EERs of 7.9\%, 5.7\%, 3.4\%, 2.7\% with 50, 100, 200, and 500 graphs, respectively, in the test sample. The test samples were formed from graphs randomly selected from all available graphs. Therefore, for each test sample, each feature may have graphs from different samples. While this type of sampling is not realistic, the results show that authentication on uncontrolled free-text with small test samples is possible. 

Mondal and Bours argue that many of the works in keystroke dynamics labeled as continuous authentication are in fact periodic authentication (PA) \cite{bours2015continuous,mondal2016combining}. Periodic refers to authentication being performed on blocks of keystrokes rather than on each individual keystroke. The authors proposed a new way to measure the effectiveness of continuous and periodic authentication systems: the average number of imposter actions (ANIA) and average number of genuine actions (ANGA). The ANIA is the average number of keystrokes before an imposter is detected while the ANGA is the average number of keystrokes before a true user is falsely rejected. Mondal and Bours developed a trust based model capable of achieving strong performance and making a decision after each keystroke. For most users, the algorithm never rejected the true user and reported an ANIA of 304 keystrokes (for the system trained with the least amount of imposter data). The authors present multiple scenarios with varying degrees of imposter data being known to the system. 

In summary, existing methods currently fail to authenticate users quickly, until our own recent work \cite{cvpr2019}, often requiring 500 or more keystrokes. Furthermore, existing algorithms treat monograph and digraph features equally instead of studying their individual impact on performance. In contrast, we show here that by using a fused matching score, calculated from the feature importance of monographs and digraphs (Section \ref{section:feature_fusion}), and our novel ITAD metric, we can achieve authentication with small numbers of keystrokes (fewer than 100). Our results are presented using detection error tradeoff (DET curves), EER, and ANGA/ANIA. In all cases, our methods are shown to outperform the state-of-the-art.

\section{Features}
\label{section:features}

Timing information recorded from keystrokes can be considered a time series. In its raw form, it is non-stationary because the time interval between keystrokes can occur at any interval and is not sampled at a continuous rate \cite{Gurdal2018ANM}. Working with non-stationary time series data can be very challenging and one of the common approaches to extract stationary data is differencing \cite{hamilton1994d,montgomery2015introduction}. The concept of differencing in keystroke dynamics may sound foreign, but in fact goes as far back as the 1980s when researchers used digraphs defined as the time taken to type two consecutive characters \cite{gaines1980authentication}. For their study, the authors only had access to the time a key was pressed down, so they believed digraphs to be the lowest level feature in their experiment. A survey of 187 papers \cite{teh2013survey}, found that 49\% used digraphs, 41\% used monographs, 5\% used pressure, and 5\% used other features (pressure is not considered in this paper as it requires special hardware to collect the data). Additionally, Teh \textit{et al.} point out that research investigating and comparing common features used for keystroke dynamics is missing. This could be very beneficial to the keystroke dynamics field by providing insight to which features are most explanative of user behavior.

The features commonly used today are the result of differencing. Monographs are defined as the time between when a key is pressed down to when it is released. Digraphs, or flight times, are usually defined in literature to be the time between two connective key-presses. In this work, four different definitions of digraphs are used, referred to as DD, UD, DU, and UU. D corresponds to a key-down event and U corresponds to a key-up event. The four digraph features are the time from the first key either being pressed (D) or released (U) to the time the second key is pressed or released. The monograph feature and the four digraph features can be seen in Fig. \ref{figure:features}. Similar to work done in \cite{huang2017benchmarking}, the graphs are only considered if their durations fall in a specific range to eliminate digraph durations that span pauses or typing sessions. 

Although trigraphs and other $n$-graphs have shown to be highly representative of users, no trigraph features or n-graph features are used in this work due to the focus on fast authentication. Using only the English alphabet, there are $26$ distinct monographs, $26 \times 26 = 676$ distinct digraphs, and $26 \times 26 \times 26 = 17,576$ distinct trigraphs. The numbers are much larger when including punctuation, numbers, and other function keys. While of course not all digraphs, trigraphs, or $n$-graphs have the same probability of occurrence, it is clear to see that with minimal keystrokes in the test sample, the probability of getting trigraphs or larger $n$-graphs that match a given profile is low.

\section{Datasets}
\label{section:datasets}

In this paper, two publicly available datasets, the Clarkson II uncontrolled free-text dataset \cite{DBLPconficb2017} and the Buffalo partially controlled free-text dataset \cite{sun2016shared}, are used to validate our results. 

The Buffalo free-text dataset consists of a total of 148 participants who contributed a combined 2.14 million keystrokes \cite{sun2016shared}. The Buffalo dataset is split into two categories: baseline and rotation. The baseline set has 75 users where the same keyboard is used. The rotation set has 73 users and three different keyboards are used. Within both rotation and baseline, there were three identical sessions consisting of transcribing Steve Job’s commencement speech at Stanford University, free-text response to survey questions, and an image, as well as some tasks designed to mimic daily work such as responding to emails and freely surfing the internet. We consider the Buffalo dataset partially controlled free-text because it is a combination of free-text and transcribing tasks. The rotation enables researchers to study the effects of different keyboards on typing behavior.

The Clarkson II dataset was collected at Clarkson University \cite{DBLPconficb2017}. Containing over 12.9 million keystrokes across 103 users, to the best of our knowledge, it is the largest free-text dataset available where an average user has 125,000 keystrokes. This dataset is different from the other publicly available datasets as all keystrokes are recorded as users interact normally with their computers. A keylogger ran in the background of participants computers, passively recording all of their keystrokes regardless of application or context. Users had the option of temporarily disabling the keylogger to protect their private information. 

As discussed in Section \ref{section:related_work}, the performance of algorithms on the Clarkson II dataset compared to other more controlled free-text datasets is always worse \cite{huang2017benchmarking}. The Buffalo dataset is partially controlled, due to containing free-text and transcribing tasks, whereas the Clarkson II dataset is completely uncontrolled. As a result, we expect the performance of all algorithms to be better on the Buffalo dataset.

In previous works, digraphs and keystrokes are both common methods of measuring the amount of data in the profile and test sample. For example, a test sample size of 50 DD digraphs contains 50 DD digraphs and all other graph features that occurred while typing those DD digraphs, and a test sample size of 50 keystrokes contains all graph features that occur within those 50 keystrokes. To best compare our results to literature, we provide a table of feature occurrence for the monograph and four digraph features for both datasets in the Appendix (Table \ref{table:feature_freq}). Table \ref{table:feature_freq} can be used to freely convert keystrokes to DD digraphs and vice versa. In this paper, our results will be presented in terms of DD digraphs with the exception of Section \ref{section:results:anga_ania} where keystrokes are used.

\section{Algorithms}
\label{section:algorithms}

Instance-based algorithms compare a single instance (occurrence) of a graph from the test sample to the profile. The test sample itself typically consists a block of graphs. In contrast, distribution-based algorithms construct a probability density function (PDF) of a graph for the profile and for the test sample. The two PDFs are then compared to determine dissimilarity or similarity. Instance-based algorithms are extensively seen in fixed-text research, but seldomly seen for free-text. Distribution-based algorithms are commonly used for free-text. However, they require more graphs in the test sample for comparisons to be made \cite{cvpr2019}. 
Profiles for a graph are only built if there are four or more occurrences of that graph (for both instance-based and distribution-based). Requiring a minimum of four graphs is based upon previous works \cite{huang2017benchmarking} as well as our own empirical observations.
The test sample, however, can be a single instance of a graph for instance-based algorithms, but four or more instances for each graph are needed for distribution-based algorithms. 

The instance-based algorithms used in this paper include Manhattan \cite{manhattan}, scaled Manhattan \cite{manhattan}, Mahalanobis \cite{maha}, transformed Mahalanobis \cite{biosig}, and the ITAD metric. The distribution-based algorithm used is the KDE algorithm introduced by Huang \textit{et al.} \cite{huang2017benchmarking}. 
We describe these algorithms as follows.

\subsection{Instance-based}
\label{section:instance_based}

The algorithms are used to compute a distance or similarity score from the keystrokes in the test sample to the keystrokes in the profile of the authorized user. 
The distance or similarity score represents how likely it was that test sample came from the authorized user. Depending on the scores, the test user is either allowed continued access or is locked out of the system.

\subsubsection{Manhattan distance}
\label{section:manhattan}

The scaled Manhattan and Manhattan distance metrics were used by Kilhourhy and Maxion for fixed-text keystroke dynamics \cite{killourhy2009comparing}. The scaled Manhattan distance is calculated as follows

\begin{equation}
\label{equation:scaled_manh}
D = \frac{1}{N} \sum_{i=1}^{N} \frac{|\mu_{g_{i}}-x_{i}|}{\sigma_{g_{i}}},
\end{equation}

\noindent where $N$ is the number of graphs shared between the test sample and the profile, $x_{i}$ is the individual test graph duration for the $i^{\text{th}}$ shared graph in the test sample, and $\mu_{g_{i}}$ and $\sigma_{g_{i}}$ are the mean and standard deviation of the $i^{\text{th}}$ graph in the profile \cite{killourhy2009comparing}. The scaled Manhattan distance formula has been altered slightly to become the average distance (by multiplying by $\frac{1}{N}$) allowing for better comparisons to be made between test samples with different numbers of graphs shared with the profile. The Manhattan and scaled Manhattan distances are identical, except the Manhattan distance is not divided by the standard deviation \cite{manhattan}.

\subsubsection{Mahalanobis distance}
\label{section:mahalanobis}

The Mahalanobis distance is similar to the scaled Manhattan distance and is given by

\begin{equation}
\label{equation:maha}
D = \frac{1}{N} \sqrt{\sum_{i=1}^{N} \frac{(\mu_{g_{i}}-x_{i})^2}{\sigma_{g_{i}}^2}},
\end{equation}

\noindent where $N$ is the number of graphs shared between the test sample and the profile, $x_{i}$ is the individual test graph duration for the $i^{\text{th}}$ shared graph in the test sample, and $\mu_{g_{i}}$ and $\sigma_{g_{i}}$ are the mean and standard deviation of the $i^{\text{th}}$ graph in the profile \cite{killourhy2009comparing,maha}. The Mahalanobis metric is also multiplied by $\frac{1}{N}$ to control for different numbers of shared graphs between the test sample and the profile. 

\subsubsection{Transformed Mahalanobis}
\label{section:tmah}

The transformed Mahalanobis distance metric is calculated with the same formula as the Mahalanobis metric. However, before computing the Mahalanobis distance, the profile distribution is transformed to a Gaussian distribution. The transformation is done through the cumulative distribution functions (CDFs), where the CDF value is held constant by mapping the $x$ value to the corresponding $x$ value for the desired distribution\cite{beasley2009rank,biosig}. Both the Manhattan and Mahalanobis distance metrics assume Gaussian data. Transforming the data to Gaussian and then using metrics designed for Gaussian distributed data is therefore a natural choice. The transformation process and distance score calculation can be combined as

\begin{equation}
\label{equation:tmah}
D = \frac{1}{N} \sum_{i=1}^{N} [Q^{-1}(CDF_{g_{i}}(x_{i}))]^{2},
\end{equation}

\noindent where $N$ is the number of graphs shared between the test sample and the profile, $Q^{-1}(\cdot)$ is the inverse Q-function [32], $CDF_{g_{i}}(\cdot)$ is the empirical cumulative distribution function of the $i^{\text{th}}$ graph in the profile, and $x_{i}$ is the individual test graph duration for the $i^{\text{th}}$ shared graph in the test sample \cite{biosig}. As with the scaled Manhattan and Mahalanobis distances, this metric is averaged for a fairer comparison between the profile and test sample.

\subsubsection{Instance-based Tail Area Density (ITAD) Metric}
\label{section:area_metric}

A new instance-based distance metric we propose in this paper is referred to as the ITAD metric. 
The ITAD metric makes no assumptions about distributions and solely relies on the tail area under the PDF, or the percentile value of the sample. The ITAD metric is calculated as follows:

\begin{equation}
\label{equation:area_metric}
s_{i} = \begin{cases}
CDF_{g_{i}}(x_{i}) & \text{if } \text{$x_{i}$ $\leq M_{g_{i}}$}\\
1 - CDF_{g_{i}}(x{_i}) & \text{if } \text{$x_{i}$ $> M_{g_{i}}$},
\end{cases}
\end{equation}

\noindent where $N$ is the number of graphs shared between the test sample and the profile, $CDF_{g_{i}}(\cdot)$ is the empirical cumulative distribution function of the $i^{\text{th}}$ graph in the profile, $M_{g_{i}}$ is the median of the $i^{\text{th}}$ graph in the profile, and $x_{i}$ is the individual test graph duration for the $i^{\text{th}}$ shared graph in the test sample. The ITAD metric is always between 0 and 0.5, and because it is a similarity score, the larger the $s$ value, the closer the sample is to the profile. The ITAD metric for $N$ singular graph durations is combined into a single similarity score $S$ as:

\begin{equation}
\label{equation:area_metric_combine}
S = \frac{1}{N} \sum_{i=1}^{N} {s_{i}}^{p}.
\end{equation}

The parameter $p$ serves as a scaling factor and can be tuned depending on the application. If $0<p<1$ then lower scores are scaled up more than higher scores and if $p>1$ then larger scores will be shifted down by a lesser amount than lower scores.

\begin{figure}[htb]
\begin{minipage}[b]{1.0\linewidth}
  \centering
  \centerline{\includegraphics[width=8.0cm]{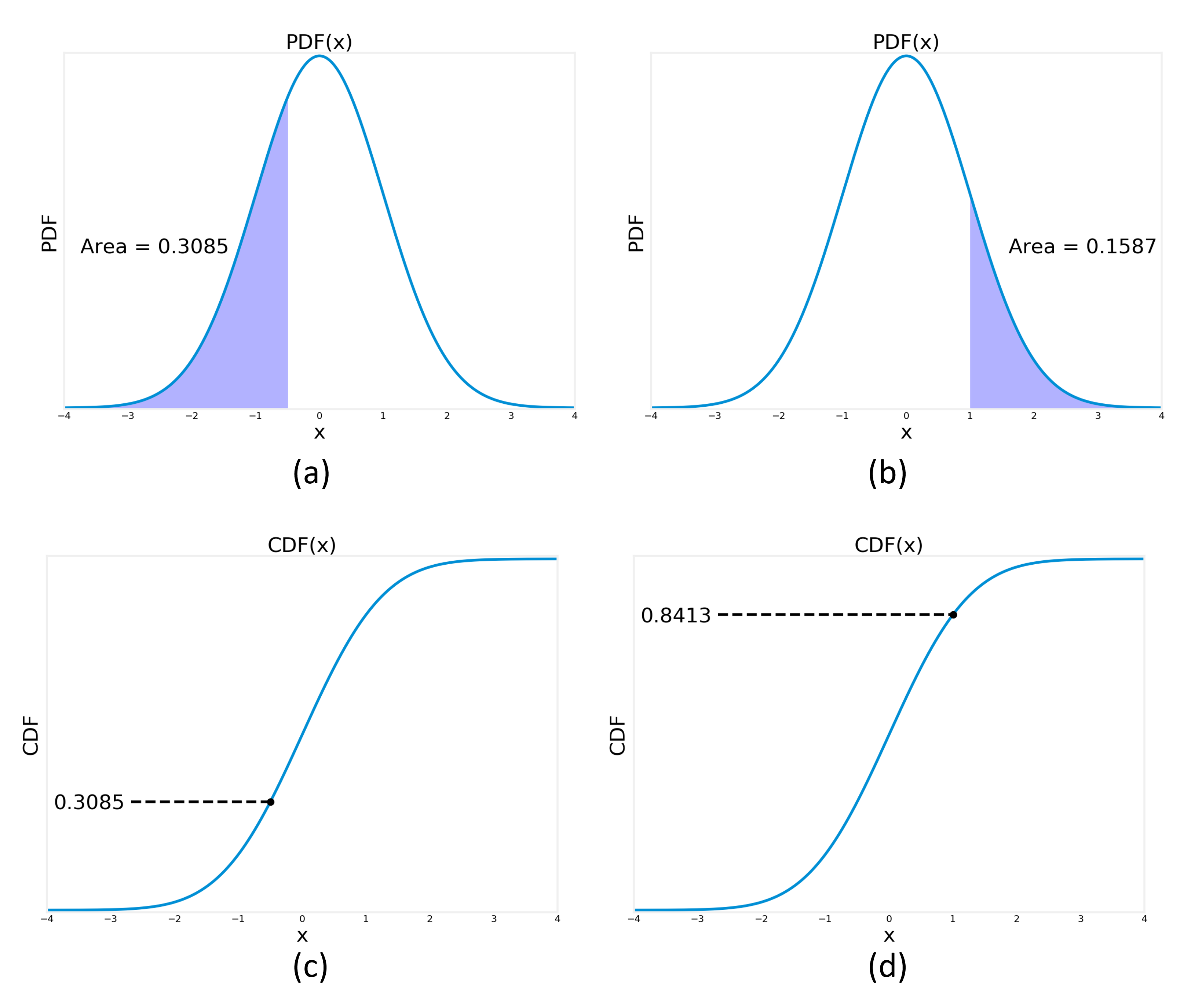}}
  \caption{Graphical representation of how the ITAD metric is computed from the PDF (a) and (b) or CDF (c) and (d).}
  \label{figure:area_metric}
\end{minipage}
\end{figure}

Figure \ref{figure:area_metric} shows a graphical representation of how the ITAD metric is computed. In terms of the PDF, the ITAD metric is computed as the tail area of the PDF. When the sample is below the median value, the ITAD metric takes the tail area on the left and when the sample is above the median, the ITAD metric takes the tail area on the right.

\begin{figure}[htb]
\begin{minipage}[b]{1.0\linewidth}
  \centering
  \centerline{\includegraphics[width=8.0cm]{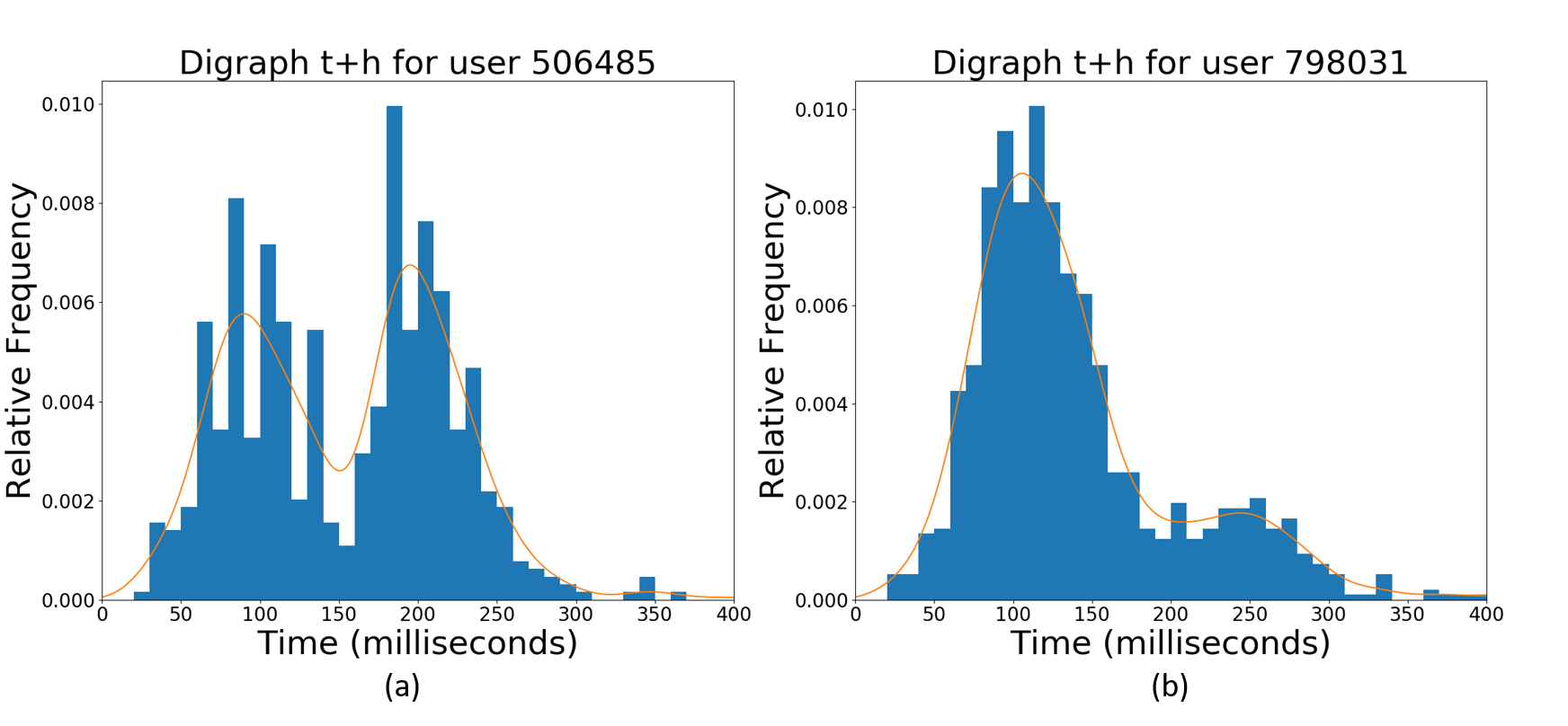}}
  \caption{Two digraph PDF's exhibiting clear non-Gaussian behavior from the Clarkson II dataset. The y-axis is relative frequency of occurrence and the x-axis is the time in milliseconds of the ``t+h'' digraph.}
  \label{figure:digraph_pdfs}
\end{minipage}
\end{figure}

Many of the previous works using instance-based algorithms have used Euclidean distance, Mahalanobis distance, or probability scores all of which were based on a mean and variance from user profiles \cite{monrose1997authentication,biosig}. These methods rely on the data having a meaningful mean and variance, which is not necessarily the case for non-Gaussian data. Figure \ref{figure:digraph_pdfs} shows the PDFs of ``t+h'' digraphs for two users which are clearly not Gaussian, suggesting a Gaussian approximation may not be the best assumption. 
For example, the sample mean of the data in Figure \ref{figure:digraph_pdfs} (a) will be directly in between the two peaks of the data. For this distribution, the distance from the mean will always be high, not because test samples are anomalous, but rather because the model does not fit the data well. The ITAD metric takes the percentile value, which causes it to be more resistant to outliers than the sample mean. 

However, the main power of the ITAD metric is from its non-parametric estimation of underlying distributions, which, as illustrated in Figure \ref{figure:digraph_pdfs}, are often not Gaussian. As a result, similarity or distance metrics such as Mahalanobis and Manhattan distance, which assume a Gaussian distribution, perform worse than our newly proposed ITAD metric. The Mahalanobis and Manhattan distances work well when each distribution is Gaussian because they provide a framework of normalizing each distribution to zero mean and unit variance, allowing for straightforward combination of multiple samples from different distributions. The ITAD metric can be thought of as providing a similar framework, but for when the distributions are not all of one type (i.e. Gaussian). The ITAD metric determines the average similarity between multiple non-parametric distributions (determined empirically from a set of previous observations) and a new observation.

\subsection{KDE}
\label{section:kde}

The distribution-based algorithm studied in this paper is the kernel density estimation (KDE) metric \cite{huang2017benchmarking,cvpr2019}. KDE is a non-parametric method used to estimate the PDF of a random variable. Here, it is used to create a PDF of the times for each graph from a finite number ($>4$) of samples \cite{silverman2018density}. KDE is distribution-based and therefore requires four or more of the same graph in both the profile and the test sample. In this paper, the python library scikit-learn's implementation of Gaussian KDE for PDF estimation \cite{scikit-learn} is used. Once the PDFs are estimated, the absolute difference between the PDFs from the profile and test samples is calculated, summed, and then averaged across all the different digraphs to produce one scalar value:

\begin{equation}
\label{eq_kde}
D = \frac{1}{N} \sum_{i=1}^{N} [PDF_{train}(x_i)-PDF_{test}(x_i)].
\end{equation}

\section{Comparison of Algorithms}
\label{section:comparing_algs}

In this section, to compare the six algorithms, an experiment is conducted using the Clarkson II and the Buffalo datasets. Profiles are built with 10,000 consecutive DD digraphs from the authorized user and tested with sample sizes ranging from 10 to 1,000 DD digraphs from the authorized user and impostors. The impostor’s data is taken from all other users. The profiles are randomly selected from all available graphs and the genuine test samples are selected from the remaining graphs. Results are averaged across users and 10 independent subsets of the data to ensure representative results. To compute a distance score for each algorithm, the profile for a graph must contain at least four or more occurrences. For the distribution-based algorithm, the test sample must also contain four or more occurrences. Instance-based algorithms, on the other hand, can compute a distance score from only a single instance of a graph in the test sample.

Table \ref{table:compare_alg_table_clarkson} shows the EERs for the 6 different algorithms using the monograph and digraph features individually and fused for the Clarkson II dataset. The score level fusion process for the monograph and digraph features is discussed in detail in Section \ref{section:feature_selection}. There are two general trends that can be seen from Table \ref{table:compare_alg_table_clarkson}. First, the performance of the ITAD metric and transformed Mahalanobis distance is similar and outperforms all other metrics for the individual features and the fused matching score. Second, the best performance for every algorithm is achieved when using the fused matching score. It can also be seen that the best performing individual features are the UU and DD digraphs followed by the DU digraph, monographs, and the UD digraph. 

The KDE algorithm has slight exceptions to the above trends. Due to being distribution-based, KDE requires more graphs in the test sample. With only 50 DD digraphs, a stable EER cannot be computed for the digraphs. As a result, the monograph feature performs best overall (by default) followed closely by the fused matching score (which essentially is just the monograph). The trends are the same with the Buffalo dataset, but it is worth noting that the EERs are slightly lower for the Buffalo dataset compared to the Clarkson II dataset, for given number of digraphs, due to the uncontrolled nature of the Clarkson II dataset. 

\newcolumntype{g}{>{\columncolor{Gray}}c}
\begin{table}[htb]
\centering
\caption{Equal Error Rates (EERs) for the six algorithms using the monograph and digraph features individually and fused for the Clarkson II dataset. The EERs are produced when there are 50 DD digraphs in the test sample. In most cases, the ITAD metric yields the lowest EERs for each feature individually, and fused. Similarly, the performance of the algorithms improve with our fused matching score.}
\tabcolsep=0.10cm
\begin{tabular}{|c|c|c|c|c|c|g|}
\hline
\multirow{2}{*}{Algorithm} & \multicolumn{6}{|c|}{Feature}\\
\cline{2-7}
& M & DD & UD & DU & UU & Fused\\
\hline
\cellcolor{Gray} ITAD & \cellcolor{Gray} 0.248 & \cellcolor{Gray} 0.197 & \cellcolor{Gray} 0.296 & \cellcolor{Gray} 0.202 & \cellcolor{Gray} 0.164 & \cellcolor{Gray} 0.123 \\
KDE & 0.220 & 0.500 & 0.500 & 0.500 & 0.500 & 0.235 \\
Manhattan & 0.290 & 0.282 & 0.329 & 0.256 & 0.243 & 0.231 \\
Scaled Manhattan & 0.244 & 0.225 & 0.301 & 0.224 & 0.211 & 0.176 \\
Mahalanobis & 0.243 & 0.278 & 0.337 & 0.268 & 0.270 & 0.223 \\
Transformed Mahalanobis & 0.252 & 0.194 & 0.277 & 0.208 & 0.161 & 0.129 \\
\hline
\end{tabular}
\label{table:compare_alg_table_clarkson}
\end{table}

\begin{figure}[htb]
\begin{minipage}[b]{1.0\linewidth}
  \centering
  \centerline{\includegraphics[width=8.0cm]{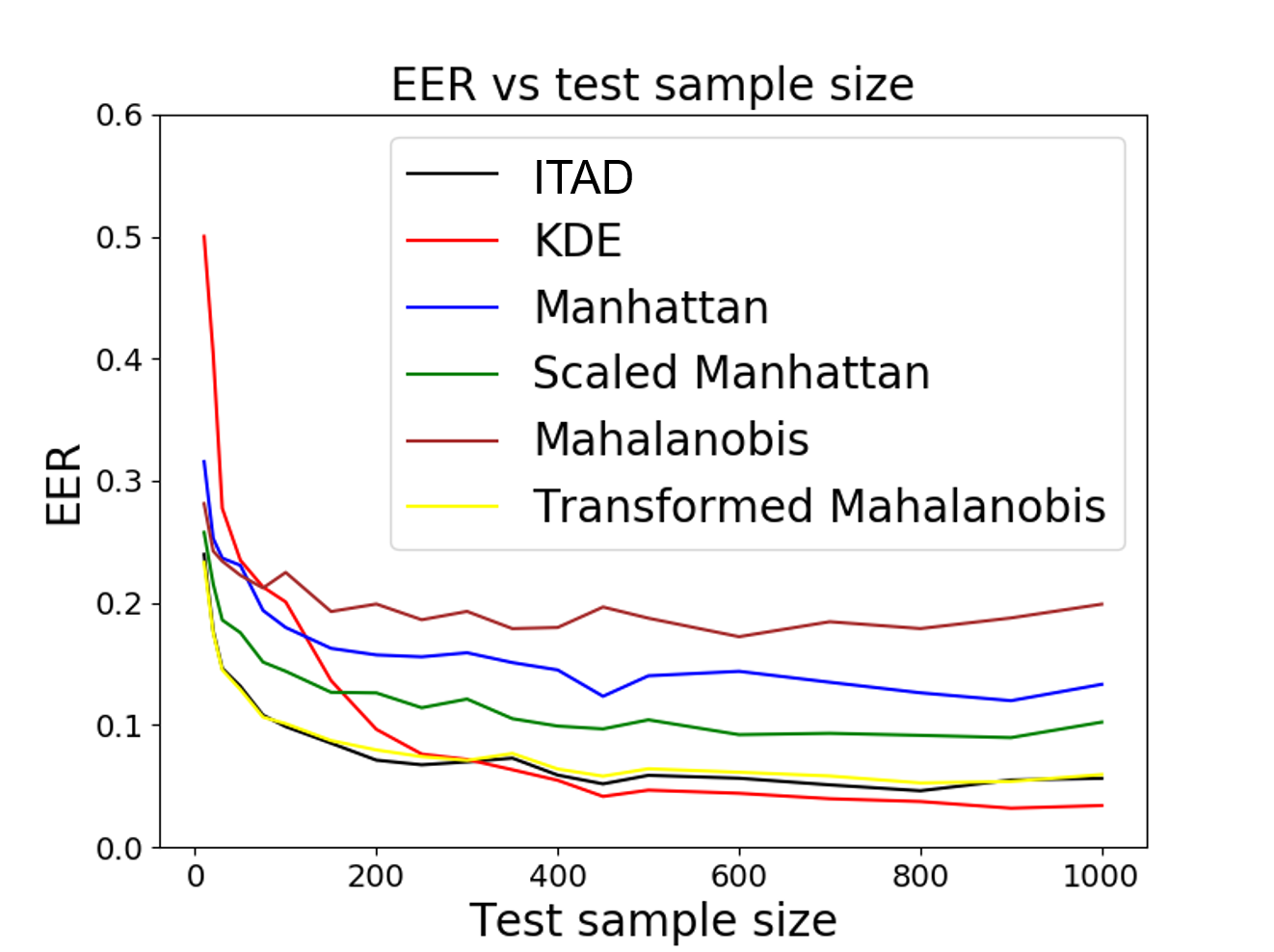}}
  \caption{EER versus test sample size (in terms of DD digraphs) for the Clarkson II dataset for 6 different algorithms. The instance-based algorithms achieve lower EERs with fewer digraphs than the distribution-based metric (KDE). With approximately 300-400 or more digraphs, the KDE metric begins to outperform the instance-based metrics.}
  \label{figure:comparing_algs_dd_clarkson}
\end{minipage}
\end{figure}

Figure \ref{figure:comparing_algs_dd_clarkson} shows the EERs for varying numbers of DD digraphs in the test sample. The instance-based algorithms outperform the distribution-based algorithm for small test sample sizes. When the number of DD digraphs is less than 80, a stable EER cannot be computed for the KDE algorithm and is set to 0.5 (random guessing). At roughly 300 DD digraphs, the distribution-based metric starts to perform better than the instance-based metrics.

Requiring 300 or more DD digraphs is far too many keystrokes for fast authentication. After 300 DD digraphs, an imposter could have typed 4-5 sentences (see Table \ref{table:intro_table}) leaving users vulnerable. Therefore, we focus on instance-based metrics and in particular, the transformed Mahalanobis and ITAD metrics. These metrics outperform the other instance-based metrics for all test sample sizes and distribution-based metrics for small test sample sizes.

\section{Feature Selection}
\label{section:feature_selection}

Many keystroke dynamics works use monographs and digraphs \cite{teh2013survey}. These features have been shown to separate users based on their typing behaviors. 
In this section, we investigate which features are important for user authentication. The feature importances are then used to determine an optimal fused matching score further improving authentication results. The feature importances are only shown for the ITAD metric, because the overall trends are the same for every algorithm.

\subsection{Feature Importance}
\label{section:feature_importance}

Feature importance is determined using a random forest classifier from both the Clarkson II uncontrolled free-text dataset \cite{DBLPconficb2017} and the partially controlled free-text Buffalo dataset \cite{sun2016shared}. The feature importance is taken as the mean decrease in impurity (MDI) from the random forest classifier. MDI is defined as the total decrease in node impurity (weighted by the probability of reaching that node) averaged over all the trees in the ensemble \cite{breiman2001random}. The probability of reaching the node is approximated by the proportion of samples reaching that node. The scikit-learn implementation of random forests and MDI are used in this paper \cite{scikit-learn}.

The feature importances are calculated for the monograph and four digraph features with 10,000 DD digraphs in the profile and using different numbers of DD digraphs in the test sample: 10, 20, 50, 100, and 200. 
A one versus all random forest classifier is built for each user. The inputs to the random forest are the scores from the ITAD metric for each feature. To ensure the the importances are representative of the data, they are calculated for each user 50 times with different subsets of user data and then averaged together. The average feature importances are reported in Tables \ref{table:feat_imp_10000} and \ref{table:feat_imp_10000_buf} for the Clarkson II and Buffalo datasets. 
\begin{table}[htb]
\centering
\caption{Relative feature importances at 10, 20, 50, 100, and 200 DD digraphs in the test sample with 10,000 DD digraphs in the profile for the Clarkson II dataset. For each test sample size, the cells highlighted in blue and red denote the feature with the highest and lowest importance.
}
\tabcolsep=0.10cm
\begin{tabular}{|c|c|c|c|c|c|}
\hline
\multirow{2}{*}{Feature} & \multicolumn{5}{|c|}{Test Sample Size}\\
\cline{2-6}
& 10 & 20 & 50 & 100 & 200\\
\hline
M & \cellcolor{max} 0.236 & \cellcolor{max} 0.241 & 0.193 & 0.157 & 0.144\\
DD & 0.221 & 0.207 & 0.229 & 0.246 & 0.240\\
UD & \cellcolor{min} 0.129 & \cellcolor{min} 0.107 & \cellcolor{min} 0.088 & \cellcolor{min} 0.091 & \cellcolor{min} 0.095\\
DU & 0.201 & 0.207 & 0.217 & 0.208 & 0.211\\
UU & 0.214 & 0.238 & \cellcolor{max} 0.273 & \cellcolor{max} 0.299 & \cellcolor{max} 0.31\\
\hline
\end{tabular}

\label{table:feat_imp_10000}
\end{table}

\begin{table}[htb]
\centering
\caption{Relative feature importances at 10, 20, 30,  50, 100, and 200 DD digraphs in the test sample with 10,000 DD digraphs in the profile for the Buffalo dataset. For each test sample size, the cells highlighted in blue and red denote the feature with the highest and lowest importance. 
}
\tabcolsep=0.10cm
\begin{tabular}{|c|c|c|c|c|c|}
\hline
\multirow{2}{*}{Feature} & \multicolumn{5}{|c|}{Test Sample Size}\\
\cline{2-6}
& 10 & 20 & 50 & 100 & 200\\
\hline
M & \cellcolor{max} 0.284 & \cellcolor{max} 0.269 & 0.228 & 0.174 & 0.140\\
DD & 0.209 & 0.221 & 0.235 & 0.256 & 0.270\\
UD & \cellcolor{min} 0.123 & \cellcolor{min} 0.101 &\cellcolor{min} 0.083 & \cellcolor{min} 0.082 & \cellcolor{min} 0.085\\
DU & 0.169 & 0.178 & 0.174 & 0.193 & 0.203\\
UU & 0.214 & 0.232 & \cellcolor{max} 0.280 & \cellcolor{max} 0.295 & \cellcolor{max} 0.302\\
\hline
\end{tabular}
\label{table:feat_imp_10000_buf}
\end{table}

The least important feature for both datasets and all test sample sizes is the UD digraph. The most important feature for both datasets is between the monograph and UU digraph. For smaller test sample sizes the monograph feature becomes more important and for larger test sample sizes the UU digraph is more important.

While the feature importances are only shown for the ITAD metric, the overall trends are the same for every algorithm. The only exception is for KDE which finds monographs far more important at smaller test sample sizes (as seen in Section \ref{section:comparing_algs}). This is because with small test samples sizes it is far more likely to see four of the same monograph than four of the same digraph (KDE is distribution-based and needs four or more occurrences in both the profile and test set to compute a distance score).

\subsection{Score-level Fusion}
\label{section:feature_fusion}

While some features are more important than others, they all contribute to the overall classification. All five features are fused at the score level to combine the individual graph features into a single matching score. A weighted average of the graph scores is taken using the feature importances calculated from the MDI as the weights. This allows the five graph scores to be combined into a single fused matching score. Thresholding is performed on this single matching score to produce DET curves and EERs. This process was used in Section \ref{section:comparing_algs} to compare the six algorithms.

\begin{figure}[htb]
\begin{minipage}[b]{1.0\linewidth}
  \centering
  \centerline{\includegraphics[width=8.0cm]{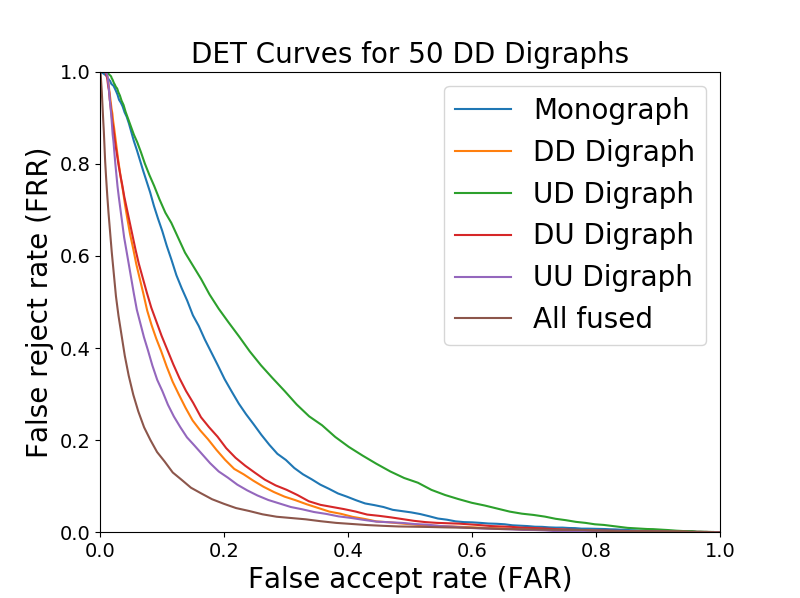}}
  \caption{DET curves for the Clarkson II dataset for the five graph features as well as the fused case for a profile size of 10,000 DD digraphs and test sample size of 50 DD digraphs. The best performing feature is the UU digraph and the worst performing feature is the UD digraph, consistent with the feature importance. A clear improvement can be seen from fusing the five features.}
  \label{figure:det_features_clarkson}
\end{minipage}
\end{figure}

Figure \ref{figure:det_features_clarkson} shows the performance of the features individually and fused using the ITAD metric for the Clarkson II dataset. The trends across the Clarkson II and Buffalo datasets are identical and therefore only the results for Clarkson II are shown. The feature importances are consistent with the DET curves. Features with higher importance have lower EERs and features with lower importance have higher EERs. It can also be clearly seen that fusing the five features together yields the best overall performance. While the EERs are shown only for the ITAD metric, just as with the feature importances, the overall trends are still the same for every algorithm. Again, the KDE algorithm is different because monographs contribute to the overall decision far more at smaller test sample sizes.

\section{Evaluation of the ITAD Metric}
\label{section:results}

The performance of the ITAD metric is evaluated with different test sample sizes, profile sizes, as well as in terms of the average number of genuine actions (ANGA) and the average number of imposter actions (ANIA). The evaluations are done in terms of DD digraphs (except for ANGA and ANIA which are presented in terms of keystrokes). In the three following subsections, the ITAD metric is used with the fusion of the five graph features to obtain results. We have used $p=\frac{1}{2}$ for combining the individual similarity scores into a single similarity score. We use DET curves, EERs, ANGA, and ANIA to present our results.

\subsection{Effect of Test Sample Size}
\label{section:results:test_size}

\begin{figure}[htb]
\begin{minipage}[b]{1.0\linewidth}
  \centering
  \centerline{\includegraphics[width=8.0cm]{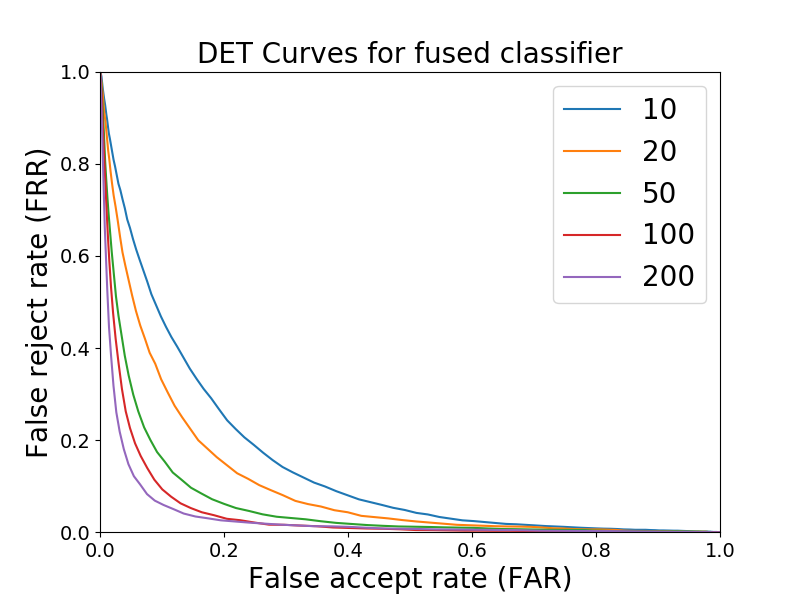}}
  \caption{DET curves for the Clarkson II dataset with the fused matching score and ITAD metric. The profile size is 10,000 DD digraphs and the test sample size ranges from 10 to 200 DD digraphs. As the test sample size increases the performance improves, but with diminishing returns.}
  \label{figure:varying_test_clarkson}
\end{minipage}
\end{figure}

\begin{figure}[htb]
\begin{minipage}[b]{1.0\linewidth}
  \centering
  \centerline{\includegraphics[width=8.0cm]{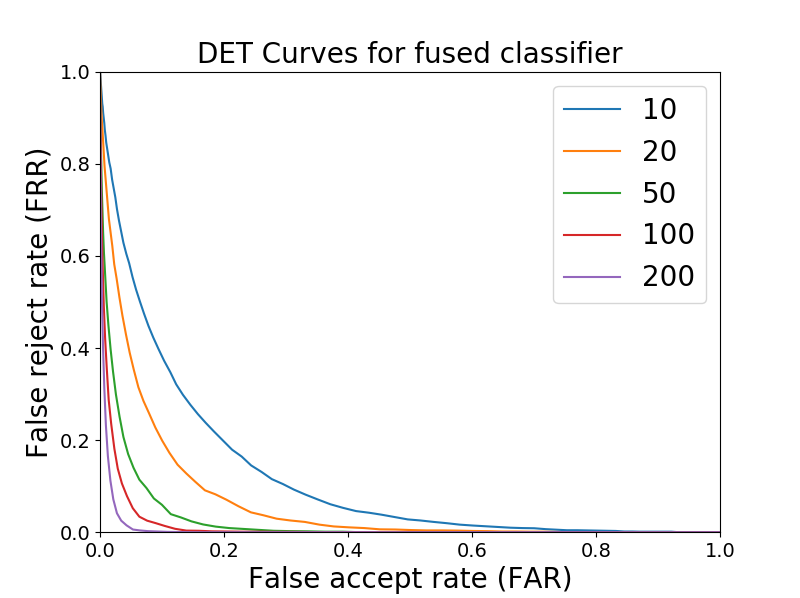}}
  \caption{DET curves for the Buffalo dataset with the fused matching score and ITAD metric. The profile size is 10,000 DD digraphs and the test sample size is varied from 10 to 200 DD digraphs. As the test sample size increases the performance improves, but with diminishing returns.}
  \label{figure:varying_test_buffalo}
\end{minipage}
\end{figure}

The performance of our algorithm is heavily dependent on the number of DD digraphs (or keystrokes) present in the test sample. With large numbers of DD digraphs in the test sample, accuracy will be better, and with fewer DD digraphs, the accuracy will be worse. Fast authentication (fewer DD digraphs) is preferable as intruders should be detected as soon as possible to mitigate risk. Figures \ref{figure:varying_test_clarkson} and \ref{figure:varying_test_buffalo} show the DET curves for varying test sample sizes for both the Clarkson II and Buffalo datasets. The curves are produced with 10,000 DD digraphs in the user profiles. Monte Carlo cross validation is performed and the experiment is repeated 50 times using different subsets of user data.

As can be seen for both the Clarkson II and Buffalo datasets, the more DD digraphs in the test sample the better the performance. This is consistent with previous works using distribution based algorithms \cite{huang2015effect}. In general, the performance on the Buffalo dataset is better than the Clarkson II dataset due to the uncontrolled nature of the Clarkson II dataset. 
The EERs for all test sample sizes and both datasets are summarized in Table \ref{table:varying_test_sizes}. The lowest EERs for the Clarkson II and Buffalo datasets are 7.8\% and 3.0\% with 200 DD digraphs in the test sample. 

Previous work achieved best EERs on the Clarkson II dataset of 35.3\% and 15.3\% for test sample sizes of 100 and 200 DD digraphs respectively \cite{cvpr2019}. The ITAD metric achieves EERs of 12.3\%, 9.7\%, 7.8\% for the Clarkson II dataset with 50, 100, and 200 DD digraphs in the test sample. For the Buffalo dataset, the ITAD metric achieves EERs of 8.0\%, 5.3\%, and 3.0\% with 50, 100, and 200 DD digraphs in the test sample. In addition to allowing authentication with fewer than 100 DD digraphs, the ITAD metric shows improvement over the existing state-of-the-art algorithms for test sample sizes fewer than 300 DD digraphs.

\begin{table}[htb]
\centering
\caption{EERs for varying test sample sizes (in terms of DD digraphs) for the Clarkson II and Buffalo datasets using the ITAD metric with the fused matching score. There are 10,000 DD digraphs used in the profile.}
\tabcolsep=0.1cm
\begin{tabular}{|c|c|c|c|c|c|}
\hline
\multirow{2}{*}{Dataset} & \multicolumn{5}{|c|}{Test Sample Size}\\
\cline{2-6}
& 10 & 20 & 50 & 100 & 200 \\ \hline
Clarkson II & 0.221 & 0.177 & 0.123 & 0.097 & 0.078 \\ \hline
Buffalo & 0.199 & 0.136 & 0.080 & 0.053 & 0.030 \\ \hline
\end{tabular}
\label{table:varying_test_sizes}
\end{table}

\subsection{Effect of Profile Size}
\label{section:results:train_size}

In this section, the performance of our algorithm is evaluated depending on the amount of DD digraphs in the profile. Building a user profile is a necessary part of keystroke dynamics authentication. While it is important imposters are detected after as few DD digraphs (or keystrokes) as possible, convenient use of the system is important as well. Users need to be able to enroll quickly (with fewer DD digraphs) or they may lose interest and decide our biometric recognition system is not worth using. According to \cite{cpm}, the average number of characters per minute is 187. This means it would take approximately one hour of continuous typing to collect 10,000 DD digraphs. 

\begin{figure}[htb]
\begin{minipage}[b]{1.0\linewidth}
  \centering
  \centerline{\includegraphics[width=8.0cm]{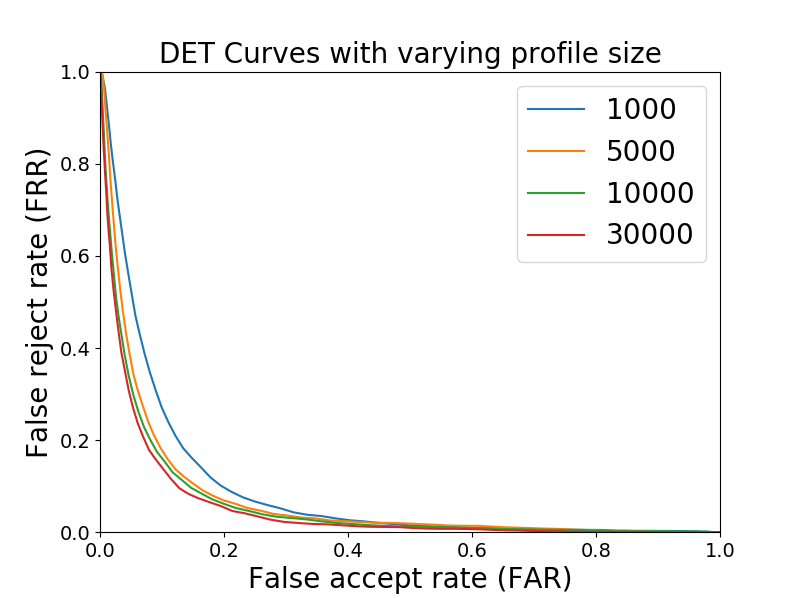}}
  \caption{DET curves for the Clarkson II dataset using the fused matching score and the ITAD metric. The profile size is varied from 1,000 to 30,000 DD digraphs with 50 DD digraphs in the test sample. As the profile size increases the performance improves, but with diminishing returns.}
  \label{figure:varying_train_clarkson}
\end{minipage}
\end{figure}

\begin{figure}[htb]
\begin{minipage}[b]{1.0\linewidth}
  \centering
  \centerline{\includegraphics[width=8.0cm]{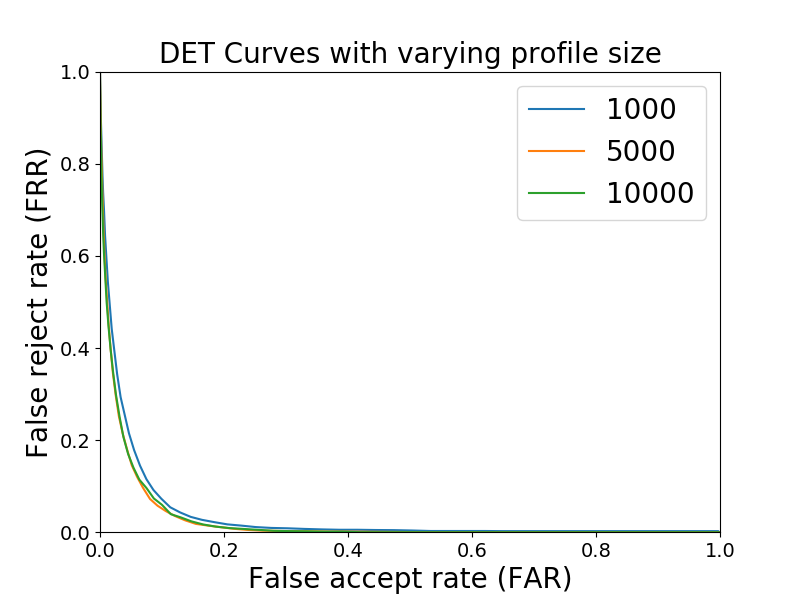}}
  \caption{DET curves for the Buffalo dataset using the fused matching score and the ITAD metric. The profile size is varied from 1,000 to 10,000 DD digraphs with 50 DD digraphs in the test sample. As the profile size increases the performance improves, but with diminishing returns.}
  \label{figure:varying_train_buffalo}
\end{minipage}
\end{figure}

Figures \ref{figure:varying_train_clarkson} and \ref{figure:varying_train_buffalo} show the DET curves for varying profile sizes for both the Clarkson II and Buffalo datasets. As the profile size increases the performance improves, but suffers from diminishing returns. It was found in \cite{huang2015effect} that the performance improved as the profile size increased (only when profile and test sizes were both larger otherwise performance decreased). We find there is no decrease in performance, but instead, performance plateaus. This is promising as an adequate user profile may be constructed with as few as 1,000 DD digraphs, allowing for fast enrollment.

\subsection{ANGA and ANIA}
\label{section:results:anga_ania}

Another way of presenting keystroke dynamics system performance is through the average number of genuine actions (ANGA) before an authorized user is rejected and the average number of imposter actions (ANIA) before an imposter is locked out \cite{bours2015continuous,mondal2016combining}. The ANGA and ANIA are directly related to the block size, FAR, and FRR as $\text{ANGA}=\text{block size} / \text{FRR}$ and $\text{ANIA}=\text{block size} / (1 - \text{FAR})$.

To compare performance in terms of ANGA and ANIA, three points along the DET curve of test sample size of 20 are selected. Point `a' favors security, point `b' weights convenience and security equally (the equal error rate), and point `c' favors convenience. The FAR, FRR, ANGA, and ANIA for the aforementioned points are shown in Table \ref{table:anga_ania} in terms of keystrokes (converted from DD digraphs using Table \ref{table:feature_freq}).

\begin{table}[htb]
\centering
\caption{Average number of Genuine Actions (ANGA) and Average Number of Imposter Actions (ANIA), in terms of keystrokes (see Table \ref{table:feature_freq}), for points on DET curves with test sample size of 20. The points ‘a’, ‘b’, and ‘c’ illustrate the tradeoff between security and convenience. These results are compared with \cite{bours2015continuous}.}
\begin{tabular}{|c|c|c|c|c|}
\hline
Algorithm & FAR & FRR & ANGA & ANIA \\
\hhline{|=|=|=|=|=|}
`a' Clarkson II & 0.0196 & 0.7552 & 40 & 32\\
`b' Clarkson II & 0.1768 & 0.1768 & 169 & 37\\
`c' Clarkson II & 0.7346 & 0.0085 & 3483 & 113\\
\hhline{|=|=|=|=|=|}
`a' Buffalo & 0.0338 & 0.4900 & 51 & 26\\
`b' Buffalo & 0.1343 & 0.1343 & 182 & 30\\
`c' Buffalo & 0.6500 & 0.0016 & 15250 & 71\\
\hhline{|=|=|=|=|=|}
Mondal and Bours & - & - & $\infty$ & 304\\
\hline
\end{tabular}
\label{table:anga_ania}
\end{table}

In order for a continuous authentication system to be useful, it cannot reject the authorized user too frequently; therefore, convenience is prioritized above security. As a result, we focus on point `c' where our algorithm achieves ANGAs and ANIAs of 3,483 and 113 keystrokes for the Clarkson II dataset and 15,250 and 71 keystrokes for the Buffalo dataset. In other words, the authorized user expects on average one false reject after 3,483 keystrokes and an imposter to be rejected after 113 keystrokes. For the Buffalo dataset, the authorized user expects on average one false reject after 15,250 keystrokes. However, an imposter would be rejected after on average 71 keystrokes. This allows for additional security to be added while only minimally impacting the authorized user. 

While our ANGA is worse than Mondal and Bours \cite{bours2015continuous} (see Table \ref{table:anga_ania})), it is still sufficiently large to have minimal impact on user experience (their experiments were conducted with uncontrolled free-text data). The ANIA, on the other hand, has been reduced significantly to more quickly detect and lockout would-be imposters robustly protecting user data. It is worth noting that we have only reported Mondal and Bours best case results which was valid only for roughly 60\% of users and the models were tuned with 50\% of the impostor users. For the other 40\% of users, the performance had lower ANGAs and/or higher ANIAs.

\section{Conclusions and Future Work}
\label{section:conclusion}

In this paper, we propose the ITAD metric, a novel instance-based algorithm, to reduce the number of DD digraphs (or keystrokes) required to authenticate users. We also investigate the effectiveness of monographs and digraphs, commonly used features in the keystroke dynamics literature, for user authentication. The most important features for fast authentication were determined to be the monograph, UU digraph, and DD digraph. However, all the features contributed information about who was typing. Scores from the five features were weighted with their feature importance to construct a new matching score providing best authentication results. Our fused matching score, when combined with the ITAD metric, is shown to outperform the state-of-the-art across two publicly available datasets allowing us to detect imposters faster and more robustly protect user data.

Previous work achieved best EERs on the Clarkson II dataset \cite{DBLPconficb2017} of 35.3\% and 15.3\% for test sample sizes of 100, and 200 DD digraphs respectively \cite{cvpr2019}. Our novel ITAD metric achieves EERs of 12.3\%, 9.7\%, and 7.8\% for the Clarkson II dataset and 8.0\%,  5.3\%, and 3.0\% for the Buffalo dataset \cite{sun2016shared} with 50, 100, and 200 DD digraphs in the test sample, a noticeable improvement over existing state-of-the-art methods. Furthermore, our algorithm achieves ANGAs of 3,483 and 15,250 keystrokes for the Clarkson II dataset and ANIAs of 113 and 71 keystrokes for the Buffalo dataset, a significant improvement over the previous best ANIA of 304 keystrokes \cite{bours2015continuous}. 
Our fused matching score, when combined with the ITAD metric, is capable of detecting imposters more than twice as fast as the previous state-of-the-art.

Future work includes investigating the effects of old or outdated profiles on performance because typing patterns have been shown to change or fluctuate over time. How often a genuine user would be rejected and its effect on usability will also be investigated to determine an appropriate balance between ANGA and ANIA (convenience and security). Other avenues of research include training machine learning and deep learning algorithms, such as neural networks, convolutional neural networks, and recurrent neural networks for keystroke dynamics.

\appendix[Feature Frequency]

In many keystroke dynamics works, digraphs and keystrokes are both common methods of measuring the amount of data in the profile and test sample. Table \ref{table:feature_freq} shows feature occurrence for the monograph and four digraph features for both datasets and can be used to freely convert keystrokes to DD digraphs and vice versa. The feature frequencies have been normalized in terms of the DD digraph and represent the number of graphs, on average, that can be expected after 1 DD digraph.

\begin{table}[htb]
\centering
\caption{
Frequencies of occurrence of the features relative to the DD digraph. 
There will be more keystrokes and monographs than digraphs because of pauses in typing. The Buffalo dataset is closer to the ideal case as the dataset is partially controlled, whereas for the uncontrolled Clarkson II dataset there are many more keystrokes than graphs.}

\begin{tabular}{|c|c|c|c|c|c|c|}
\hline
Dataset & Keystrokes & M & DD & UD & DU & UU\\
\hline
Clarkson II\cite{DBLPconficb2017} & 1.48 & 1.36 & 1.00 & 0.95 & 1.08 & 0.98\\
Buffalo \cite{sun2016shared} & 1.22 & 1.16 & 1.00 & 0.95 & 1.06 & 1.00\\
\hline
\end{tabular}

\label{table:feature_freq}
\end{table}

Table \ref{table:feature_freq} shows, for both datasets, there are more keystrokes than monographs and more monographs than digraphs. This is the result of filtering out keystrokes that likely occur across sessions and other outliers, e.g. typing one-handed while eating a sandwich. 
In fact, previous work on the Clarkson II dataset has shown that by filtering ``gibberish'' keystrokes from the keystroke data, performance improves \cite{huang2016effects}. By removing the ``gibberish'' keystrokes, the ratio of keystrokes to usable graphs becomes closer to 1. In this paper, no ``gibberish'' filtering is done to ensure our results are most representative of typical user typing behavior. 

It should be noted that under ideal conditions, keystrokes and monographs should have the same frequency of occurrence. 
However, digraphs will always be slightly fewer because if there are $N$ keystrokes there will be $N$ monographs and $N-1$ digraphs. For example, if the average typing session lasts 10 keystrokes before pausing then only 90\% of keystrokes will result in digraphs. 

\ifCLASSOPTIONcompsoc
  \section*{Acknowledgments}
\else
  \section*{Acknowledgment}
\fi

This work is supported in part by the NSF CPS award 1646542, Clarkson Niklas Ignite Fellowship, and material is based upon work supported by the Center for Identification Technology Research (CITeR) and the NSF under Grants 1650503 and 1314792.

\ifCLASSOPTIONcaptionsoff
  \newpage
\fi

\bibliographystyle{IEEEtran}
\bibliography{refs}

\begin{IEEEbiography}[
{\includegraphics[width=1in,height=1.25in,clip,keepaspectratio]{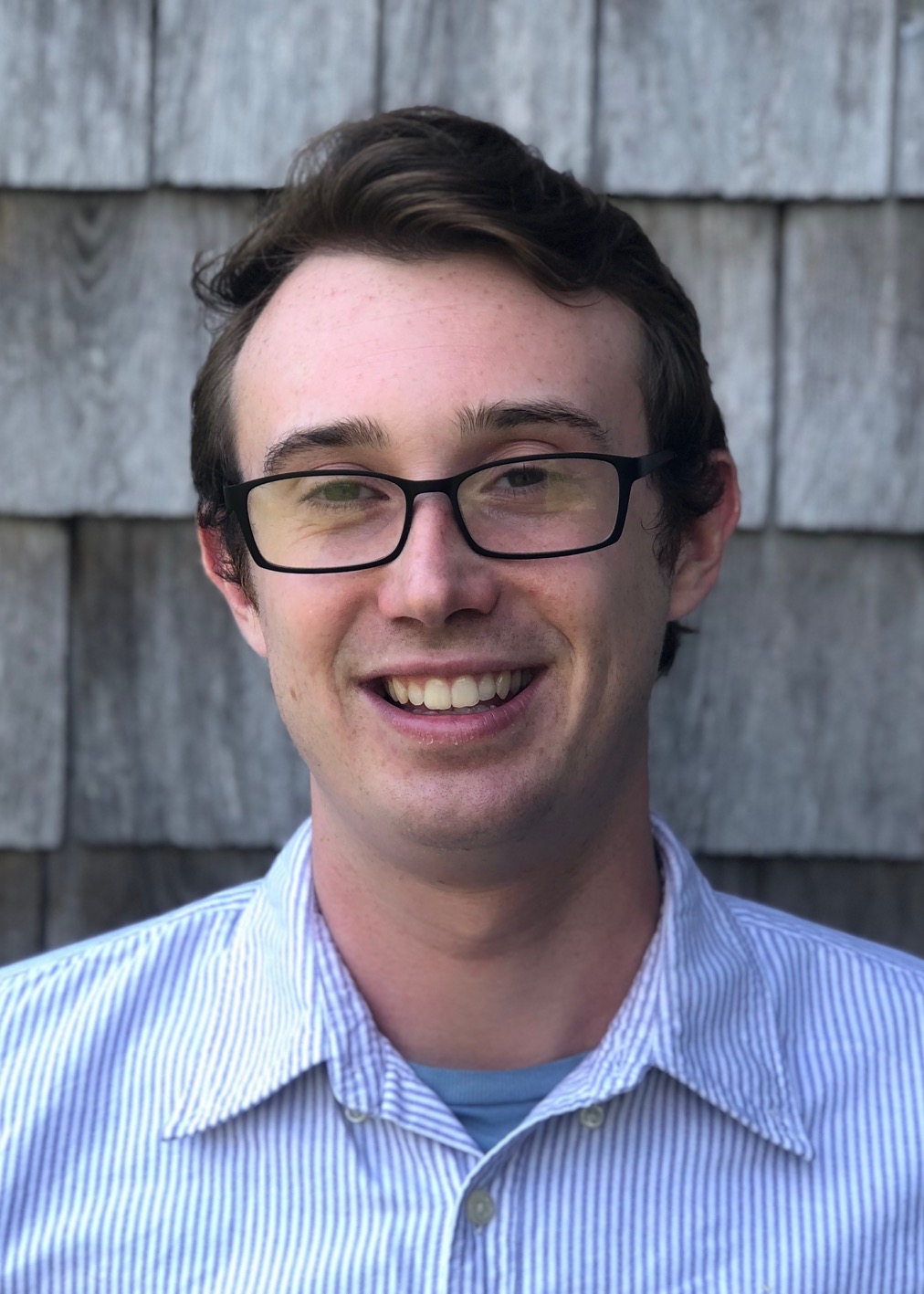}}]{Blaine Ayotte}
received the B.S. degree in mathematics and physics from St. Lawrence University, Canton, New York, USA in 2017, an M.S. degree in electrical engineering at Clarkson University, Potsdam, NY, USA, and is currently pursuing a Ph.D. in electrical engineering at Clarkson University, Potsdam, NY, USA. His research interest includes behavioral biometrics such as keystroke dynamics and authentication on mobile devices. Mr. Ayotte’s awards and honors include Niklas Ignite Research Fellowship.
\end{IEEEbiography}

\vskip -2\baselineskip plus -1fil

\begin{IEEEbiography}[{\includegraphics[width=1in,height=1.25in,clip,keepaspectratio]{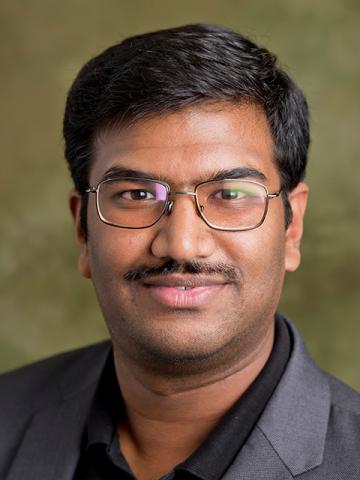}}]{Mahesh Banavar}
(S’08–M’11–SM’17) received the B.E. degree in telecommunications engineering from Visvesvaraya Technological University, India, in 2005, and the M.S. and Ph.D. degrees in electrical engineering from Arizona State University, Tempe, in 2007 and 2010, respectively. He is currently an Associate Professor with the Department of Electrical and Computer Engineering, Clarkson University, Potsdam, NY, USA. His interests include node localization, detection and estimation algorithms, and user-behavior-based cybersecurity applications. 
\end{IEEEbiography}

\vskip -2\baselineskip plus -1fil

\begin{IEEEbiography}[{\includegraphics[width=1in,height=1.25in,clip,keepaspectratio]{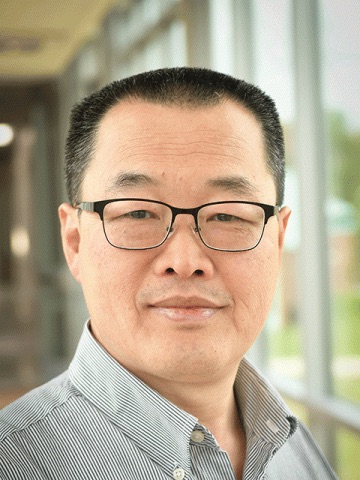}}]{Daqing Hou}
is Professor and Director of Software Engineering in the Department of Electrical and Computer Engineering at Clarkson University. He received his doctoral degree in Computing Science from The University of Alberta. His research interests include software engineering and applications, behavioral biometrics, and engineering education. His work is funded by various agencies, including the National Science Foundation, Department of Defense, Air Force Research Lab, IBM, Facebook, among others. He has published over 70 peer-reviewed research papers.
\end{IEEEbiography}

\vskip -2\baselineskip plus -1fil

\begin{IEEEbiography}[{\includegraphics[width=1in,height=1.25in,clip,keepaspectratio]{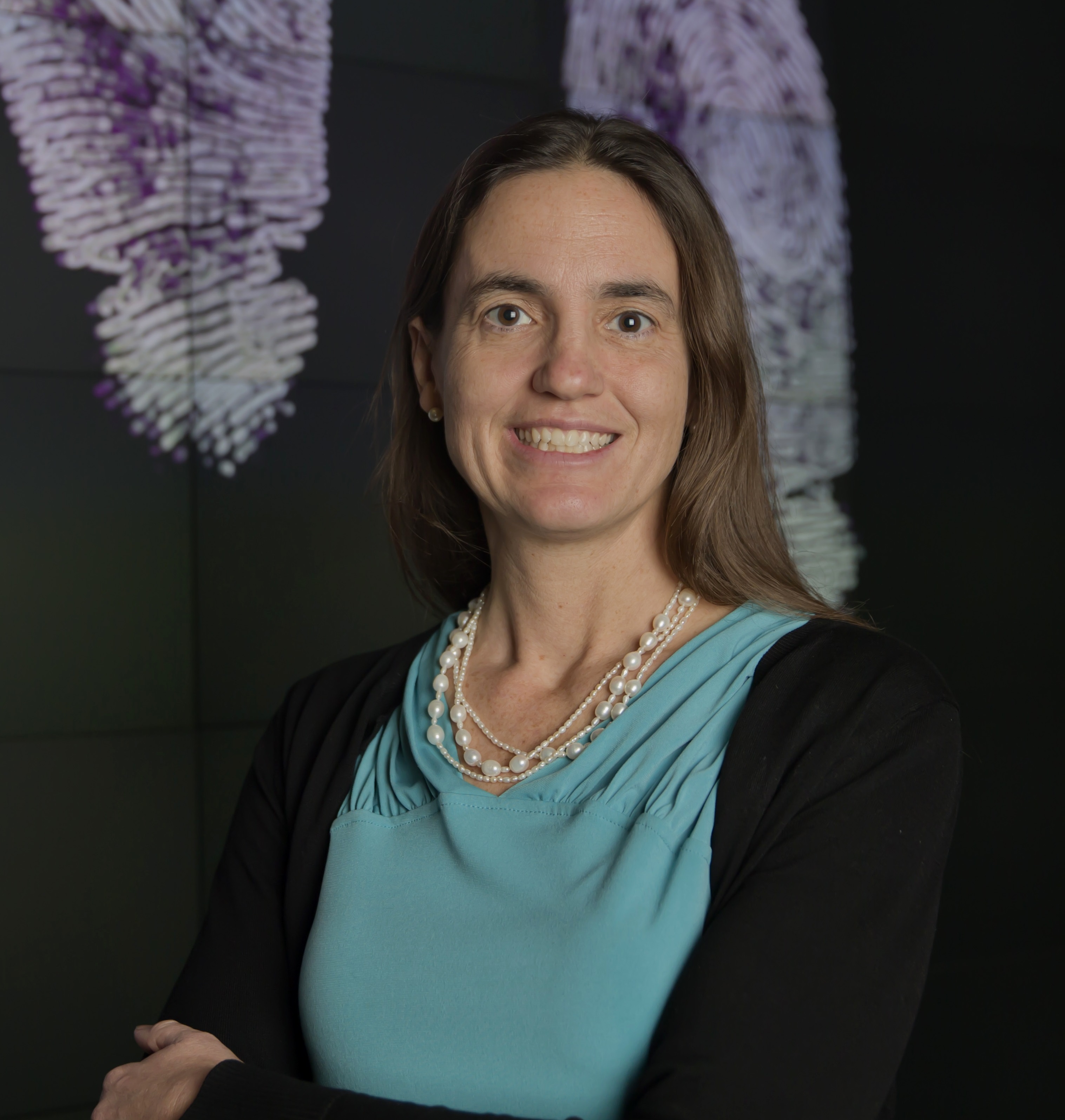}}]{Stephanie Schuckers}
is the Paynter-Krigman Endowed Professor in Engineering Science in the Department of Electrical and Computer Engineering at Clarkson University and serves as the Director of the Center for Identification Technology Research (CITeR), a National Science Foundation Industry/University Cooperative Research Center. She received her doctoral degree in Electrical Engineering from The University of Michigan. Professor Schuckers research focuses on processing and interpreting signals which arise from the human body.  Her work is funded from various sources, including National Science Foundation, Department of Homeland Security, and private industry, among others. She has started her own business, testified for US Congress, and has over 40 journal publications as well as over 60 other academic publications.
\end{IEEEbiography}

\vfill

\enlargethispage{-5in}

\end{document}